\documentclass[10pt]{article}
\usepackage{graphicx}
\usepackage{url}

\def\Title#1{\begin{center} {\Large #1 } \end{center}}
\def\Author#1{\begin{center}{ \sc #1} \end{center}}
\def\Address#1{\begin{center}{ \it #1} \end{center}}

\newcommand\pubblock{\rightline{\begin{tabular}{l} Proceedings of the Second Annual LHCP\\ 
         \pubdate  \end{tabular}}}

\newenvironment{Abstract}{\begin{quotation} \begin{center} 
             \large ABSTRACT \end{center}\bigskip 
      \begin{center}\begin{large}}{\end{large}\end{center} \end{quotation}}

\newenvironment{Presented}{\begin{quotation} \begin{center} 
             PRESENTED AT\end{center}\bigskip 
      \begin{center}\begin{large}}{\end{large}\end{center} \end{quotation}}

\def\Acknowledgements{\bigskip  \bigskip \begin{center} \begin{large}
             \bf ACKNOWLEDGEMENTS \end{large}\end{center}}

\usepackage{color}

\newcommand\pubdate{\today}

\def\affiliation{
Blackett Lab., Imperial College, London \\
and \\
Particle Physics, Denys Wilkinson Building, Oxford }

\begin{document}

\large
\begin{titlepage}
\pubblock

\vfill
\Title{Statistical Issues in Searches for New Physics }
\vfill

\Author{ Louis Lyons }
\Address{\affiliation}
\vfill
\begin{Abstract}

Given the cost, both financial and even more importantly in terms of human effort, in 
building High Energy Physics accelerators and detectors and running them, it is important 
to use good statistical techniques in analysing data. This talk covers some of the 
statistical issues that arise in searches for New Physics. They include topics such as: $\hfill$
\begin{itemize}
\item{Should we insist on the 5 sigma criterion for discovery claims?}
\item{$P(A|B)$ is not the same as $P(B|A)$.}
\item{The meaning of $p$-values}.
\item{What is Wilks Theorem and when does it not apply?}
\item{How should we deal with the `Look Elsewhere Effect'?}
\item{Dealing with systematics such as background parametrisation.}
\item{Coverage: What is it and does my method have the correct coverage?}
\item{The use of $p_0$ v $p_1$ plots.}
\end{itemize}
\end{Abstract}
\vfill

\begin{Presented}
The Second Annual Conference\\
 on Large Hadron Collider Physics \\
Columbia University, New York, U.S.A \\ 
June 2-7, 2014
\end{Presented}
\vfill
\end{titlepage}
\def\thefootnote{\fnsymbol{footnote}}
\setcounter{footnote}{0}
%

\normalsize 








\section{Theme}
We consider the situation where data are used to distinguish between 
two possible theories - the null hypothesis $H_0$ that the Standard
Model is all that is needed, and the alternative $H_1$ that there is 
also evidence for some exciting New Physics in our data. 
We assume that a data statistic $t$ is defined and that probability density functions
$f_0(t)$ and $f_1(t)$ under the two hypotheses are know 
(subject perhaps to the values of nuisance parameters being defined).
We discuss very briefly several different statistical issues that arise.

Accelerators and detectors are expensive, both in terms of money and 
human effort. It is thus important to invest effort in performing a
good statistical analysis of the data, in order to extract the best
information from it.

\section{Why 5$\sigma$  for Discovery?}
Statisticians ridicule the insistence of achieving a $p$-value  at 
least as small as $3\times 10^{-7}$ (equivalent to a significance of 5$\sigma$) before 
claiming a discovery. They say that people do not know probability
distributions out in such extreme tails; this is especially true for
systematic effects. 

The $`5\sigma$ standard' is supposed to provide protection against false
discovery claims from the following effects:
\begin{itemize}
\item{History: There are many cases of $3\sigma$ and $4\sigma$ effects that 
have disappeared with more data.}
\item{LEE:
This is discussed in Section \ref{LEE}.}
\item{Systematics: These are usually more difficult to estimate than statistical
uncertainties. An analysis that is dominated by systematic effects which are 
overestimated by a factor of 2 and which claims an apparent $6\sigma$ 
discovery in reality has only a much less interesting 
$3\sigma$ effect.}   
\item{Subconscious Bayes factor:
Even when an analysis does not use explicit Bayesian techniques, Physicists 
subsciously tend to assess the Bayesian probabilities $p(H_i|t)$ of 
$H_0$ and $H_1$ in deciding which hypothesis to accept:
\begin{equation}
\frac{p(H_1|t)}{p(H_0|t)} = \frac{p(t|H_1)}{p(t|H_0)} 
\frac{\pi(H_1)}{\pi(H_0)}   
\end{equation}   
Here $t$ is the data statistic; the first ratio on the right-hand side is 
the likelihood ratio; and the second 
is the ratio of prior probabilities for the hypotheses. If $H_1$ 
involves something very unexpected (e.g. neutrinos travel faster than the 
speed of light; energy is not conserved in LHC collisions; etc), $\pi_1/\pi_0$
will be very small, and so the likelihood ratio would need to be extremely large,
in order to have a convincing $p(H_1|t)$. This is the basis for the oft-quoted 
`Extraordinary claims require extraordinary evidence'. }
\end{itemize}

The last three items above clearly vary from  one analysis to another. Thus 
it is unreasonable to have a single criterion (5$\sigma$) for all experiments.
We might well require a higher level of significance for a claim to have discovered 
gravitational waves or sterile neutrinos than for the expected production of
single top-quarks at a hadron collider.
Ref. \cite{five_sigma} is an attempt to stimulate discussion of having 
different criteria for different analyses.

\section{$P(A|B) \ne P(B|A)$}
It is worth reminding your Laboratory or University media contact personnel that 
a very small  probability $P(data|theory)$ for getting your data, according to
some theory, does not imply that the probablity $P(theory|data)$
of the theory, given the data, is also very small.
Thus in an experiment to measure the speed of neutrinos, if the probability of 
getting the observed timing, assuming that neutrinos don't travel faster than light,
is very small, it is incorrect to assume that this implies almost certainty
that neutrinos travel faster than the speed of light.  

If anyone still believes that $P(A|B) = P(B|A)$, remind them that the 
probability of being pregnant, given that the person is female, is $\sim$3\%,
while the probability of being female, given that they are pregnant, is 
considerably larger.  

\section{$p$-values}
For a given hypothesis $H$,  the pdf $f(t|H)$ is the
probability or probability density of observing $t$, the chosen data statistic.
For a simple counting experiment, $t$ might be just the number of observed 
events. In more complicated cases, it could be a likelihood ratio. The $p$-value
is the probability of a value of $t$ at least as large as the observed value\footnote{We
are assuming that interesting deviations from $H_0$ would involve an 
{\bf increase} in $t$.}. Small
$p$-values imply that our data and $H$ are incompatible; this could be because the
theory is incorrect, the modelling of the effects of our detector on $f(t|H)$ is
inadequate, 
we have a very large statistical fluctuation, etc.

$p$-values tend to be misunderstood. It is crucial to remember that they do {\bf not} 
give the probability of the theory being true. A typical demonstration of this 
misunderstanding is the jibe directed against Particle Physicists that `they do not know
what they are doing because half of their exclusions based on $p < 5\%$ turn out to be 
wrong.' The fallacy of such reasoning is demonstrated by imagining a series of 1000 measurements 
designed to test energy conservation at the LHC. Assuming that energy really is conserved, with
a cut at $5\%$, we expect about 20 of these measurements to reject the hypothesis of energy 
conservation, and all of them will be `wrong'. Provided you understand what $p$-values
are, you will not find this paradoxical. 

Statisticians\cite{statisticians} also tend to attack $p$-values
in that numerically they can be very much smaller than likelihood ratios.
However, we should not expect them to be similar because 
$p$-values refer to the tail area of the data statistic $t$ for a single 
hypothesis, while the likelihood ratio is the relative heights for $t$ 
of the pdf's for two hypotheses. Also the criticism is rather like 
complaining that, in comparing elephants and mice, their mass ratio is 
too extreme compared with the ratio of their heights.

\section{Wilks Theorem}
We assume that we are comparing some data (e.g. a mass histogram) with two theories 
$H_0$ and $H_1$. If $H_0$ is true, we expect $\Delta S = S_0 - S_1$ to be small or negative;
here $S_i$ is the weighted sum of squares for the comparison of $H_i$ with the data. 
Wilks Theorem states that under certain circumstances, $\Delta S$ will 
be distributed as $\chi^2$ with $\nu_0 - \nu_1$ degrees of freedom ($\nu_i$ is the
ndf for the fit of hypothesis $H_i$ to the data). This is useful in helping us decide 
which hypothesis we prefer, in that we know the expected distribution of $\Delta S$, 
assuming $H_0$ is true, and hence we do not have to do elaborate simulations to determine its 
distribution.

The Table illustrates three different scenarios, with the last column stating whether
or not Wilks Theorem applies, with asymptotic data. The conditions for the theorem to apply are:
\begin{itemize}
\item{$H_0$ is true.}
\item{The hypotheses are nested i.e. it is possible to reduce $H_1$ to $H_0$ by a suitable
choice of the free parameters in $H_1$.}
\item{The values of the free parameters required to achieve this are all defined, and
not at the boudary of their range.}  
\item{The data is asymptotic.}
\end{itemize}

\begin{table}
\begin{center}
\begin{tabular} {|c|c|c|c|c|c|}
\hline   Data                &       $H_0$                  &          $H_1$              &  Nested?  &   Params OK? & W. Th. applies? \\ \hline
\hline Mass histogram & Polynomial of degree 3 & Polynomial of degree 5    &  Yes   &   Yes    &   Yes    \\
\hline Mass histogram &   Background distribution   &    Bgd + signal      &  Yes   &   No     & No    \\
\hline $\nu$ oscillation data  & Normal $\nu$ mass hierarchy   &   Inverted hierarchy & No  &  N/A    & No  \\
\hline
\end{tabular}  
\caption{Applicability of Wilks Theorem. In comparing data with two hypotheses, this depends
on whether the hypotheses are nested, whether the parameter values required to 
reduce $H_1$ to $H_0$ are all defined and not on their physical boundaries, and 
whether there is sufficient data to be in the asymptotic regime.} 
\end{center}
\label{table:Wilks}
\end{table}          

\section{Look Elsewhere Effect (LEE)}
\label{LEE}
Last month I was travelling on the London underground, and bumped into a colleague I hadn't seen for
ages. What a big coincidence that was! Well, it would have been if I had wondered in advance whether 
I would by chance meet him that day. But there were plenty of ex-colleagues I could have bumped into,
and it didn't have to be that particular day, so the overall probability 
of such an event happening by chance is much larger than I might have thought.

That is the essence of the LEE. If I observe a peak at a particular mass in a specific 
spectrum, the probability by chance of observing such an effect or larger at that position
in that spectrum is the 
local $p$-value. But the much larger chance of this happening anywhere is the global $p$-value;
their ratio is the LEE factor.

A problem is that definition of `anywhere' is imprecise. For the graduate student performing
this analysis, `elsewhere' is at any relevant mass value in that histogram (or perhaps in 
any histogram used in that analysis). But the Director General of CERN might be concerned to
avoid claiming the discovery of new effects that were in fact simply due to statistical fluctuations
in any CERN experiment, and so his `elsewhere' would be much wider than the graduate student's.

Because of this ambiguity, it is important when quoting a global $p$-value to specify your definition of `elsewhere'.

\section{Background systematics}

In a typical search, there are many possible sources of systematics that 
need to be considered. Here we discuss just one of them.

In fitting a mass distribution by the null hypothesis (background only)
or the alternative (background plus signal), it is necessary to find a
way of describing the background, for example by a specific functional form with 
free parameters. But perhaps the chosen functional form is inadequate,
and hence there is a systematic associated with the choice of function.
Ways of coping with this have included:
\begin{itemize}
\item{Try different functional forms, and for assessing the systematic, ignore
those that have a goodness of fit significantly worse than the best choice. But 
a problem is `What constitutes worse?'}
\item{Use a background subtraction method}
\item{Use a Baysian approach}
\item{Use a non-parametric method}
\item{etc.}  
\end{itemize}

A new idea is to try various functional forms, and to plot  as a function of the 
parameter of interest (e.g. the signal strength) the 
log-likelihood $LL$ for each of them, with 
possible offsets for different numbers of free parameters. Then a modified
$LL'$ is defined as the envelope of all the individual $LL$s. It is this 
widened $LL'$ that is used to make statements on the signal strength, which
incorporate the uncertainty resulting from the various functional forms.
It is a method for discrete choices that corresponds to profile likelihoods
used for continuous nuisance parameters. It has been used in the CMS 
$H^0\rightarrow \gamma \gamma$ analysis\cite{bgd_syst}.

\section{Coverage}

Consider analysing some data to obtain either a range or an upper limit for 
a parameter (e.g. the rate at which some hypothesised new particle is produced).
If this procedure was repeated many times, statistical fluctuations would result 
in differences among the determined ranges.
The fraction of these ranges that include the true value for the parameter is called the  
`coverage'. Ideally the coverage should be independent of the true value of
the parameter, and it should equal the nominal value; for supposed 1$\sigma$
intervals it should be 68$\%$. A technique which has coverage below the nominal value is serious for 
Frequentists; quoted ranges for the parameter are less likely to contain the true value
than is naively expected.

In an interesting note, Heinrich has plotted the coverage for a Poisson counting 
experiment where the intervals for the Poisson parameter are determined from the likelihood by the
$\Delta\ln L = 0.5$ rule. The plot of coverage against the Poisson mean is 
dramatically different from naive expectation (see the figure on page 10 of ref. \cite{Heinrich}).

It is important to realise that coverage is a property of the {\bf statistical procedure} 
used to extract the parameter's range, and does {\bf not} apply to your {\bf actual measurment}.

\section{$p_0$ v $ p_1$ plots}
A recent preprint\cite{Demortier} advocates the use of plots of $p_0$ versus $p_1$
for understanding various issues in comparing data with two hypothesestwo hypotheses. These include
\begin{itemize}
\item{the $CL_s$ method for excluding $H_1$;}
\item{the Punzi definition of sensitivity;}
\item{the relationship between $p$-values and likelihoods;}
\item{the probability of misleading 
evidence;} 
\item{the Law of the Iterated Logarithm; and }
\item{the Jeffreys-Lindley paradox.} 
\end{itemize}

\section{Conclusions}
In performing statistical analyses, it is important to be aware of resources 
that are available. Thus there are books written by Particle Physicists\cite{books}, 
and a useful summary of Statistics is provided by the Particle Data Group\cite{PDG}. 
Also the large collaborations have Statistics Committees, some of which 
have public web-pages\cite{web_pages}.

On the software side, ROOSTATS\cite{RooStats} is set up to deal 
with a wide range of statistical problems.

So before reinventing the wheel for your data analysis, see if Statisticians (or Particle Physicists) have
already provided a solution to your problem. In particular, do not use your own 
square wheel if a circular one already exists.

Best of luck with your analyses.

\Acknowledgements
I am grateful to members of the CDF and CMS for many lively and useful discussions.


\begin{thebibliography}{99}


\bibitem{five_sigma} L. Lyons, `Discovering the significance of $5\sigma$',
\url{http://arxiv.org/pdf/1310.1284v1.pdf} (2013).
\bibitem{statisticians} Many Bayesian statisticans, public communications (1763-2014). 
\bibitem{bgd_syst} CMS Collaboration, `Observation of the diphoton decay of 
the Higgs boson and measurement of its properties',
\url{http://arxiv.org/abs/arXiv:1407.0558} (2014). 
\bibitem{Heinrich} J. Heinrich, CDF/MEMO/STATISTICS/PUBLIC/6438, `Coverage of Error 
Bars for Poisson data',
\url{http://www-cdf.fnal.gov/physics/statistics/notes/cdf6438_coverage.pdf} (2003).
\bibitem{Demortier} L. Demortier and L. Lyons, `Testing Hypotheses in Particle Physics: Plots of 
$p_0$ v $p_1$', 
\url{http://arxiv-web3.library.cornell.edu/pdf/1408.6123v1.pdf} (August2014.
\bibitem{books} R.J. Barlow, `Statistics' (Wiley,1989); \\
O. Behnke et al. (editors), `Data Analysis in High Energy Physics: a Practical
Guide to Statistical Methods' (Wiley. 2013); \\
G. Cowan, `Statistical Data Analysis' (OUP, 1998);   \\
F. James, `Statistical Methods in Experimental Physics', (World Scientific, 2006); \\
L. Lyons, `Statistics for Nuclear and Particle Physicists' (CUP, 1986); \\
B. Roe, `Probability and Statistics in Experimental Physics' (Springer, 1992).
\bibitem{PDG} G. Cowan, `Statistics' in ``Review of Particle Properties'', \\
\url{http://pdg.lbl.gov/2013/reviews/rpp2013-rev-statistics.pdf}
\bibitem{web_pages} For example, CDF Statistics Committee, 
\url{http://www-cdf.fnal.gov/physics/statistics/statistics_home.html} \\
CMS Statistics Committee, 
\url{https://twiki.cern.ch/twiki/bin/view/CMS/StatisticsCommittee}
\bibitem{RooStats} L. Moneta et al., PoS ACAT2010 057 (2010), arXiv:1009.1003 [physics.data-an]. 


  

\end{thebibliography}
\end{document}